# On the Permutation Distribution of Independence Tests

Ehab F. Abd-Elfattah

Ain Shams University, Cairo, Egypt.

ehab@ASUnet.shams.edu.eg

**Abstract**

One of the most popular class of tests for independence between two random variables is the general class of rank statistics which are invariant under permutations. This class contains Spearman's coefficient of rank correlation statistic, Fisher-Yates statistic, weighted Mann statistic and others. Under the null hypothesis of independence these test statistics have a permutation distribution that usually the normal asymptotic theory used to approximate the p-values for these tests. In this note we suggest using a saddlepoint approach that almost exact and need no extensive simulation calculations to calculate the p-value of such class of tests.

Some key words: Independence tests; Linear rank test; Permutation distribution; Saddlepoint approximation.

## 1 Introduction

When the factors being studied are not treatments that the investigator can assign to his subjects but conditions or attributes which are inseparably attached to these subjects, an assumption that need to be tested is that an association exists between two factors in a population of subjects. Let us observe $N$ independent pairs of random variables $(X_1, Y_1)$, $(X_2, Y_2)$, ..., $(X_N, Y_N)$ and we wish to test the null hypothesis $H_0$ that the two variables $X_i$ and $Y_i$ are independent for each $i$.

Now rearrange all $N$ pairs of observations according to the magnitude of their first coordinate into the sequence $(X_{d_1}, Y_{d_1})$, $(X_{d_2}, Y_{d_2})$, ..., $(X_{d_N}, Y_{d_N})$ in such a way that $X_{d_1} < X_{d_2} < \cdots < X_{d_N}$. Then put $R_i$ equal to the rank of $Y_{d_i}$ among the observations $Y_{d_1}, Y_{d_2}, ..., Y_{d_N}$. Under the assumption of independence and assuming no ties, all $N!$ orderings $(R_1, ..., R_N)$ are equally likely with probability $1/N!$. If we willing to assume that the two factors have a positive associations, the $\{R_i\}$ should reveal an upward trend, with large values tending to occur on the right of the sequence and low values on the left. An appropriate test statistic that reflects this idea is

$$D = \sum_{i=1}^{N}(R_i - i)^2 \qquad (1)$$

with small values of $D$ indicating significance.

The statistic $D$ is related to the well known Spearman's coefficient of rank correlation statistic, $S_p$, with the relation $S_p = 1 - 6D/N(N^2 - 1)$, see Gibbons and Chakraborti (2003). It is also related to the weighted Mann statistic, $D'$, by $D' = \frac{1}{6}N(N^2 - 1) - \frac{1}{2}D$.

Expanding (1), $D$ can be written as

$$D = \frac{1}{3}N(N+1)(2N+1) - 2\sum_{i=1}^{N} iR_i$$

which gives an equivalent simple statistic

$$V' = \sum_{i=1}^{N} iR_i \qquad (2)$$

Hajek, Sidak and Sen (1999).

The statistic $V'$ is equivalent to a general class of rank statistics whose null distributions are invariant under permutations, this class can be written as

$$S = \sum_{i=1}^{N} f_N(i) f_N(R_i) \qquad (3)$$

which contains the Fisher-Yates normal score test with $f_N(i) = EU_N^{(i)}$, where $U_N^{(1)} < U_N^{(2)} < \cdots < U_N^{(N)}$ being an ordered sample of $N$ observations from the standardized normal distribution, the van der Waerden test statistic, with $f_N(i) = \Phi^{-1}(\frac{i}{N+1})$, where $\Phi$ is the standard normal distribution function and the quadrant test statistic with $f_N(i) = sign(i - \frac{N+1}{2})$.

Saddlepoint approximation to randomization distributions were introduced by Daniels (1958) and further developed by Robinson (1982) and Davison and Hinkly (1988). Booth and Buter (1990) showed that various randomization and resampling distributions are the same as certain conditional distributions and that the double saddlepoint approximation attains accuracy comparable to the single saddlepoint approach. Recently, Abd-Elfattah and Butler (2007) used the double saddlepoint approximation to calculate the p-values and confidence interval for the class of linear rank two sample statistics for censored data.

In this note we present a simple, fast and accurate saddlepoint approach that does not need any extensive permutation simulations, to calculate the exact p-value for the previous class of tests using double saddlepoint approximation. To use the double saddlepoint approximation, the following lemma reformulate the class (3) to more appropriate simple form.

**Lemma 1** *The class of statistics (3) can be written in an equivalent form as*

$$V = L^T \sum_{i=1}^{N} f_N(i) Z_i \qquad (4)$$

*where $L^T = (f_N(1), f_N(2), ..., f_N(N))$, and $Z_1, Z_2, ..., Z_N$ are $N \times 1$ vectors of the form $Z_{R_i} = \eta_i$, $i = 1, ..., N$, where the $N \times N$ identity matrix $I_N = (\eta_1, \eta_2, ..., \eta_N)$.*

**Proof.** Simple algebra. ■

For example, if $R_1 = 2$ is arithmetical rank so that $Z_2 = \eta_1$ and $\sum_{i=1}^{N} i Z_i$ has a 2 in its first component for $R_1$.

Section 2 presents the saddlepoint approximation approach. A real data example has illustrated in section 3 along with a simulation study to show the performance of the saddlepoint method. The application of the saddlepoint method to Cuzick (1982) test statistic in case of interval censoring is discussed in section 4.

## 2  Saddlepoint Approximation for Tests of Independence

Under the null hypothesis $H_0$ of independence, the permutation distribution of $V$ places a uniform distribution on the set of $N \times 1$ indicator vectors $\{Z_i\}$. This distribution may constructed from a corresponding set of i.i.d. $N \times 1$ vectors of $Multinomial(1, \theta_1, \theta_2, ..., \theta_N)$ indicators $\zeta_1, \zeta_2, ..., \zeta_N$. The permutation distribution over all one way design for which $\sum_{i=1}^{N} Z_i = (1, ..., 1)^T$ is constructed from the i.i.d. Multinomial variables as the conditional distribution

$$Z_1, ..., Z_N \stackrel{D}{=} \zeta_1, ..., \zeta_N | \sum_{i=1}^{N} \zeta_i = (1, ..., 1)^T_{N \times 1}$$

the dependence in the statistic can be removed by using $(N-1) \times 1$ vectors $Z_i^-$ and $\zeta_i^-$, the first $N-1$ components in $Z_i$ and $\zeta_i$, then

$$Z_1^-, ..., Z_n^- \stackrel{D}{=} \zeta_1^-, ..., \zeta_n^- | \sum_{i=1}^{n} \zeta_i^- = (1, ..., 1)^T_{(N-1) \times 1}$$

and then $V$ can be represented in terms of $\{Z_i^-\}$ as

$$V = L_-^T \sum_{i=1}^{N} f_N(i) Z_i^- + Q$$

where $L_-^T = (f_N(1) - f_N(N), ..., f_N(N-1) - f_N(N))$ and $Q = f_N(N) \sum_{i=1}^{N} f_N(i)$.

If $v_0$ is the observed statistic value of $V$, then the null distribution of $V$ is

$$\Pr\{V \geq v_0\} = \Pr\{T(\zeta^-) = L_-^T \sum_{i=1}^{N} f_N(i) \zeta_i^- + Q \geq v_0 \mid \sum_{i=1}^{N} \zeta_i^- = (1, ..., 1)^T\}$$

Assuming any probability vector $\{\theta_1, \theta_2, ..., \theta_N\}$ for the Multinomial distribution, the conditional distribution of $T(\zeta_1^-, \zeta_2^-, ..., \zeta_N^-)$ is the required permutation distribution which can be approximated by using the double saddlepoint approximation of Skovgaard (1987).

The p-value is approximated from the double saddlepoint procedure in which uses the joint cumulant generating function for $(T(\zeta_1^-, \zeta_2^-, ..., \zeta_N^-), \sum_{i=1}^{N} \zeta_i^-)$ given by $K(s,t) = \log M(s,t)$ where

$$M(s,t) = \prod_{i=1}^{N} \left\{ \sum_{j=1}^{N-1} \theta_j \exp(s_j + r_{ij}t) + \theta_N \right\}$$

with $s = (s_1, ..., s_{N-1})$ and $r_{ij} = f_N(i)(f_N(j) - f_N(N))$, and then

$$\Pr(V \geq v_0) \simeq 1 - \Phi(\hat{w}) - \phi(\hat{w})\left(\frac{1}{\hat{w}} - \frac{1}{\hat{u}}\right)$$

where

$$\hat{w} = \text{sgn}(\hat{t})\sqrt{2\left[-\{K(\hat{s},\hat{t}) - \hat{s}^T 1_- - v_0\hat{t}\}\right]}$$
$$\hat{u} = \hat{t}\sqrt{|K''(\hat{s},\hat{t})|/|K''_{ss}(0,0)|}.$$

and $1_-$ is $(N-1) \times 1$ vector of ones. In these expressions, $K''$ is the $N \times N$ Hessian matrix and $K''_{ss}$ is the $\partial^2/\partial s \partial s^T$ portion at $(0,0)$. The saddlepoint $(\hat{s}, \hat{t})$ solves

$$K'_{sj}(\hat{s}, \hat{t}) = \sum_{i=1}^{N} \frac{\exp(\hat{s}_j + r_{ij}\hat{t})}{\left\{\sum_{l=1}^{N-1} \exp(\hat{s}_l + r_{il}\hat{t}) + 1\right\}} = 1, \quad j = 1, ..., N-1$$

$$K'_t(\hat{s}, \hat{t}) = \sum_{i=1}^{N} \frac{\sum_{j=1}^{N-1} r_{ij} \exp(\hat{s}_j + r_{ij}\hat{t})}{\left\{\sum_{l=1}^{N-1} \exp(\hat{s}_l + r_{il}\hat{t}) + 1\right\}} + Q = v_0$$

using $\theta_i = 1/N$ the denominator saddlepoint equations have an explicit solution as $\hat{s}_0 = 0$ and this simplifies the calculations.

# 3 Example and Simulation Study

Nayak (1988) gives the failure times of transmission ($X$) and of transmission pumps ($Y$) on 15 caterpillar tractors as shown in table 1.

| X | 1641 | 5556 | 5421 | 3168 | 1534 | 6367 | 9460 | 6679 |
|---|------|------|------|------|------|------|------|------|
|   | 6142 | 5995 | 3953 | 6922 | 4210 | 5161 | 4732 |      |
| Y | 850  | 1607 | 2225 | 3223 | 3379 | 3832 | 3871 | 4142 |
|   | 4300 | 4789 | 6310 | 6311 | 6378 | 6449 | 6949 |      |

Table 1. Failure times of transmissions by Nayak (1988).

To test the independence of failure times of $X$ and $Y$, the test statistic (2) are used with $L = (1, ..., N)$, and $Q = L_N \sum_{i=1}^{N} R_i = N^2(N+1)/2$. The true (simulated) p-value was calculated by using $10^6$ permutations of the computed test statistic. The simulated p-value is then the proportion of such generations exceeding the observed statistic plus the proportion of those equal. The p-value of the saddlepoint approach is compared to the normal p-value calculated using the test statistic $(v' - E(v'))/\sqrt{Var(v')}$. The true p-value and the saddlepoint approximated p-value were 0.2768 and 0.2763, respectively, while the normal $p$-value was 0.2693.

A small simulation study has carried out to assist the performance of the saddlepoint method. Consider the general model of dependence

$$X_i = X'_i + \lambda e_i, \qquad Y_i = Y'_i + \lambda e_i, \qquad i = 1, ..., N$$

where all the variables $X'_i, Y'_i$ and $e_i$ are mutually independent and their distributions do not depend on $i$, and $\lambda$ is a real non-negative parameter. In this model the null hypothesis $H_0$ of independence is equivalent to $\lambda = 0$, whereas for $\lambda > 0$ the variables $X_i$ and $Y_i$ are dependent. Data sets are generated from this model using Logistic, Extreme value and Uniform distributions for $X'_i, Y'_i$ and $e_i$ respectively. For each value of $\lambda = 0.0, 0.5$ and sample sizes $(10, 20, 30)$, a 1000 data sets are generated and the true, saddlepoint and normal p-values are calculated using the test statistic (2). Table 2 shows the proportion of the 1000 data sets that saddlepoint p-value was closer to the true p-value than the normal p-value "Sad. Prop.", "Abs. Err. Sad." is the average absolute error of the saddlepoint p-value from the true p-value, and "Rel. Abs. Err. Sad." is the average relative absolute error of the saddlepoint p-value from the true p-value.

| $n$ | $\lambda$ | Sad. Prop. | Abs. Err. Sad. | Rel. Abs. Err. Sad |
|---|---|---|---|---|
| 10 | 0.0 | 0.944 | 0.0010 | 0.0048 |
|    | 0.5 | 0.945 | 0.0010 | 0.0050 |
| 20 | 0.0 | 0.957 | 0.0003 | 0.0016 |
|    | 0.5 | 0.956 | 0.0003 | 0.0018 |
| 30 | 0.0 | 0.943 | 0.0003 | 0.0012 |
|    | 0.5 | 0.932 | 0.0003 | 0.0013 |

Table 2. Performance under simulation for the dependence test.

The saddlepoint p-value was more accurate in 94.6% of the overall cases as compared to the normal approximation. The average absolute saddlepoint error was less than $10^{-3}$ with average relative error typically less than 0.1%. An important consideration in these saddlepoint computations is the difficulty in solving $N$ saddlepoint equations. This becomes increasingly difficult with large $N$.

## 4 Discussion

The problem of testing the independence between two variables under random censoring has taken attention of many authors, see O'Brien (1978), Wei (1980), Oakes (1982), and Gieser and Randles (1997). When one of the two variables is under interval censoring, say the first, and the second random variable is observed, Cuzick (1982) presents a linear log-rank test statistic to test the independence of two vectors in the form $\sum_{i=1}^{N} \xi_i R_{2i}$, where $\{\xi_i\}$ are given scores and $\{R_{2i}\}$ are the ranks of the observed values of the second random variable. In the linear form (4), taking $L = (\xi_1, \xi_2, ..., \xi_N)$ and $f_N(R_i) = R_{2i}$, the saddlepoint method is simply applicable. For example, Cuzick gives a survival times for 20 patients for the analysis of the relation between hemoglobin at presentation and survival in some medical clinice. The normal p-values using his asymptotic approach was 0.0505 while the true p-value and the saddlepoint p-value are 0.0516 and 0.0512, respectively.